\documentstyle[prl,aps,twocolumn]{revtex}

\begin{document}
\draft
\title{A Geometric Diagram of Separable States}
\author{Ping-Xing Chen$^{1,2\thanks{%
E-mail address: pxchen@nudt.edu.cn}}$and Cheng-Zu Li$^1$}
\address{$^1$Key Laboratory of Quantum Information, University of Science and\\
Technology of China, Chinese Academy of Sciences, Hefei 230026, People's\\
Republic of China.  \\
$^2$Department of Applied Physics, National University of Defense\\
Technology, Changsha, 410073, People's Republic of China.}
\maketitle

\begin{abstract}
This paper present a geometric diagram of a separable state: If a mixed
state $\sigma $ is separable, there are $2^{nS(\sigma )}$ linearly
independant product vectors which span the same Hilbert space as the $%
2^{nS(\sigma )}$ ``likely'' strings of $\sigma ^{\otimes n}$ do. This
diagram results in a criterion for separability which is strictly stronger
than the inorder criterion in [M.A. Nielsen and J. Kempe, Phys. Rev. Lett.
86, 5184 (2001)]. This means that the number of product bases of states of a
system has close link to the nonlocality of the system.
\end{abstract}


Quantum entanglement is one of quantum properties with no classical
counterpart. It is not only closely connected to fundamental question of
quantum mechanics, such as non-locality \cite{4} and the experimental tests
of Bell theorem \cite{5}, but also to many remarkable applications, such as
error correcting code\cite{1}, dense coding\cite{2} and teleportation\cite{3}%
. Although properties of the entanglement of bipartite pure states are now
clear to us, in realistic physical situation we should deal with mixed
states \cite{benn}, in which pure states entanglement has been degraded by
noises. A bipartite mixed state $\rho _{AB},$ owned by $A$ and $B,$ has
infinite kinds of pure states decompositions \cite{woot}, such as

\begin{equation}
\rho _{AB}=\sum_ip_i\left| \Psi _i\right\rangle \left\langle \Psi _i\right|
\label{n1}
\end{equation}
where $\sum_ip_i=1,$ $\left| \Psi _i\right\rangle $ is a pure state. If $%
\rho _{AB}$ has at least one kind of pure states decomposition where each
pure state $\left| \Psi _i\right\rangle $ is a product state, we say $\rho
_{AB}$ is separable, otherwise $\rho _{AB}$ is entangled.

It is in general very difficult to know whether a given mixed state $\rho
_{AB}$ is entangled or separable. A.Peres \cite{pere} demonstrated a simple
and practical necessary condition for separability which is called positive
partial transpose (PPT) condition. Horodecki et al \cite{horo} presented a
sufficient and necessary condition of separability of a mixed state by
positive maps on operators, and proved that the PPT criterion is also
sufficient for the bipartite systems of $2\otimes 2$ and $2\otimes 3.$
Later, there are many attentions in this direction, e.g., Ref. \cite
{6,7,8,duan,chen,9} and references therein.

Recently, Nielsen and kempe \cite{10} presented another necessary condition,
i.e., {\it inorder criterion}: the decreasingly ordered vector of the
eigenvalues for $\rho _{AB}$ is majorized by reduced matrix $\rho _A$ or $%
\rho _B.$ The inorder criterion can be expressed in the formula: 
\begin{equation}
\lambda (\rho _{AB})\prec \lambda (\rho _A),\lambda (\rho _B),  \label{n2}
\end{equation}
where $\lambda (\rho )\equiv [\lambda _1,\lambda _2,\cdots ],\lambda
_1>\lambda _2>\cdots .$ $\lambda _1,\lambda _2,\cdots $ are eigenvalues of
matrix $\rho ,$ $\rho _{A,B}=Tr_{A,B}(\rho _{AB}).$ We define $\lambda
(.)\prec \lambda ^{\prime }(.)$ if

\begin{equation}
\sum_{i=1}^k\lambda _i\leq \sum_{i=1}^k\lambda _i^{\prime }
\end{equation}
for $k=1,2,\cdots $ . The condition in Eq.(\ref{n2}) is strictly stronger
than the entropic criterion \cite{6,7}, which can be expressed as: if $\rho
_{AB}$ is separable, the global and local entropy satisfy that 
\begin{equation}
S(\rho _A),S(\rho _B)\leq S(\rho _{AB}),  \label{n3}
\end{equation}
where $S(.)$ is the von Neumann entropy. The criterions for separability Eq.(%
\ref{n2}) and (\ref{n3}) are interesting because they show that{\it \
separable states are more disordered globally than locally.}

Information entropy describes not only the disorder of the system but also
the loss of classical information \cite{11}. A mixed state denoted by Eq.(%
\ref{n1}) may be imagined to be part of an extended system which is in the
pure state

\begin{equation}
\left| \Psi _{MAB}\right\rangle =\sum_i\sqrt{p_i}\left| \Psi _i\right\rangle
_{AB}\left| m_i\right\rangle  \label{n4}
\end{equation}
or in the mixed state

\begin{equation}
\rho _{ABM}=\sum_ip_i\left| \Psi _i\right\rangle _{AB}\left\langle \Psi
_i\right| \otimes \left| m_i\right\rangle \left\langle m_i\right| ,
\label{n5}
\end{equation}
where $\left| m_i\right\rangle s$ are orthogonal states of the memory as
shown in Ref.\cite{12}. If we have no access to the memory system, we trace
over it to obtain the mixed state in Eq.(\ref{n1}). When the system became
the mixed state in Eq.(\ref{n1}) from that in Eq.(\ref{n5}), the loss of
classical information \cite{11} (the quantum mutual information between the
memory and $AB$ systems is defined as classical information $I_c$) is $%
\triangle I_c=S(\rho _{AB})$ owing to the lack of access to the memory. The
loss of the classical information may result in the loss of the entanglement
of the system. In this sense, the information entropy of a system has some
links to the entanglement of the system (e.g,. see \cite{13}). Thus the
information entropy should be relative to the separability of a state, i.e.,
there may be criterion for separability related to the information entropy
(CSRIE). To get a CSRIE, one should consider the other properties of the
state besides to the information entropy. For example, the inorder criterion
considered local inorder (or local entropy). We will show that the number of
product vectors of a state is better than the local inorder for one to get a
CRSIE.

On the other hand, consider a mixed entangled state $\sigma ,$

\begin{equation}
\sigma =\sum_{i=1}^m\lambda _i\left| \Phi _i\right\rangle \left\langle \Phi
_i\right| ,\quad \sum_{i=1}^m\lambda _i=1,  \label{a}
\end{equation}
where $\left| \Phi _i\right\rangle s$ are the eigenstates of $\sigma $ with
nonzero eigenvalues $\lambda _is.$ As shown in the paper by Bennett et al 
\cite{benn} that the $n$ copies of $\sigma ,\sigma ^{\otimes n},$ is a
mixture of $2^{nS(\sigma )}$ orthogonal ``likely'' strings of pure states.
In each of ``likely'' strings there are $\lambda _in$ copies whose states
are $\left| \Phi _i\right\rangle .$ The probability that each ``likely''
string occurs is $p=\prod_{i=1}^m\lambda _i^{n\lambda _i}.$ So the number of
orthogonal ``likely'' strings of pure states also quantify the global
inorder ( or information entropy of $AB$ systems). In this paper, we will
present a criterion for separability by comparing the number of orthogonal
likely'' strings of pure states of $\sigma ^{\otimes n}$ and the number of
product vectors of $\sigma ^{\otimes n}.$ Interesting, our criterion is
strictly stronger than the inorder one in \cite{10}. These conclusions, as
well as Ref \cite{q,f} which showed the number of product vectors of states
affects the local distinguishability of the states, means that the number of
product vectors of a mixed state are relative to the nonlocality.

Theorem 1: If a mixed state $\sigma $ is separable, then the $2^{nS(\sigma
)} $ ``likely'' strings of $\sigma ^{\otimes n}$ are vectors of a Hilbert
space $H$ spanned by $2^{nS(\sigma )}$ linearly independent product vectors,
i.e., the $2^{nS(\sigma )}$ linearly independent product vectors and the $%
2^{nS(\sigma )}$ ``likely'' strings of $\sigma ^{\otimes n}$ span the same
Hilbert space.

Proof: Suppose that the mixed state $\sigma $ is separable, then $\sigma
^{\otimes n}$ is separable. According to the criterion of pure states
decompositions of a mixed state \cite{woot,niel}, each ``likely'' strings of 
$\sigma ^{\otimes n}$, noted as $\left| string_j\right\rangle ,$ can be
expressed as a supposition of product states as follows: 
\begin{equation}
\sqrt{p}\left| string_j\right\rangle =u_{jl}\left| ps_l\right\rangle ,\qquad
j=1,\cdots ,2^{nS(\sigma )};l=1,\cdots ,M  \label{th1}
\end{equation}
$M$ is not less than $2^{nS(\sigma )}.$ $u_{jl}$ is an element of unitary
matrix $u.$ $\left| ps_l\right\rangle s,$ unnormalized, is a set of product
states decomposition of $\sigma ^{\otimes n}$. Since unitary matrix $u$ has
its invertible matrix, each product state $\left| ps_l\right\rangle $ is a
linear composition of the $2^{nS(\sigma )}$ ``likely'' strings $\left|
string_j\right\rangle s.$ The basic theory of linear algebra means that the
number of linearly independent states of $M$ product states $\left|
ps_l\right\rangle s$ is $2^{nS(\sigma )}.$ So Eq.(\ref{th1}) can be
rewritten as: 
\begin{equation}
\sqrt{p}\left| string_j\right\rangle =A_{jl^{\prime }}\left| ps_{l^{\prime
}}\right\rangle ,\qquad j,l^{\prime }=1,\cdots ,2^{nS(\sigma )}  \label{th2}
\end{equation}
where $A_{jl^{\prime }}$ is an element of matrix $A.$ $A$ is not necessarily
unitary, $\left| ps_{l^{\prime }}\right\rangle s$ are $2^{nS(\sigma )}$
linearly independent product states of $M$ product states $\left|
ps_l\right\rangle s.$ Eq. (\ref{th2}) means that each ``likely'' strings $%
\left| string_j\right\rangle $ is a linear composition of the $2^{nS(\sigma
)}$ $\left| ps_{l^{\prime }}\right\rangle s.$ So the $2^{nS(\sigma )}$
``likely'' strings $\left| string_j\right\rangle s$ and the $2^{nS(\sigma )}$
product states $\left| ps_{l^{\prime }}\right\rangle s$ span the same
Hilbert space $H$. This ends the proof.

For any mixed state $\sigma ,$ one can always find a set of linearly
independent product bases (LIPB) so that each of the $2^{nS(\sigma )}$
``likely'' strings $\left| string_j\right\rangle s$ of $\sigma ^{\otimes n}$
is a linear composition of these LIPBs only if the number of LIPBs is big
enough. This can be said that there is a space spanned by the LIPBs so that
all ``likely'' strings $\left| string_j\right\rangle s$ of $\sigma ^{\otimes
n}$ are in this space. We define this space as {\it product bases space }%
(PBS). The dimensions of the PBS must be not less than $2^{nS(\sigma )}$.
Theorem 1 implies that if $\sigma $ is a separable state there is a PBS with
the least number of dimensions, $2^{nS(\sigma )}$, all ``likely'' strings $%
\left| string_j\right\rangle s$ of $\sigma ^{\otimes n}$ belong to the least
dimensional PBS (LPBS).

Theorem 2: The product bases space criterion for separability, Theorem 1, is
strictly stronger than the disorder criterion in Eq.(\ref{n2}) and the
entropic criterion in Eq.(\ref{n3}).

Proof: Let's first prove that the entropic criterion is equivalent to the
disorder criterion at the infinite copies of a mixed state limit. All
eigenvalues of $\sigma _A^{\otimes n}$ are equal when $n\rightarrow \infty ,$
and similarly for $\sigma _B^{\otimes n}$ and $\sigma _{AB}^{\otimes n}.$ If 
$S(\sigma _A),S(\sigma _B)\leq S(\sigma _{AB}),$ then $S(\sigma _A^{\otimes
n}),S(\sigma _B^{\otimes n})\leq S(\sigma _{AB}^{\otimes n})$. This implies $%
\lambda (\sigma _A^{\otimes n}),\lambda (\sigma _B^{\otimes n})\succ \lambda
(\sigma _{AB}^{\otimes n}).$ The converse is obvious as shown in the Ref. 
\cite{10}.

Let's turn to prove the theorem2. Suppose that the dimensions of LPBS of $%
\sigma ^{\otimes n}$ is $M$, i.e., each ``likely'' strings $\left|
string_j\right\rangle $ is a linear composition of the $M$ product states,
then the number of nonzero eigenvalues of $\sigma _A^{\otimes n}$ or $\sigma
_B^{\otimes n}$ is at most $M.$ This follows that: 
\begin{equation}
S(\sigma _A^{\otimes n}),S(\sigma _B^{\otimes n})\leq \ln M  \label{th3}
\end{equation}
If $\sigma $ is a separable state Theorem 1 follows that

\begin{equation}
2^{nS(\sigma )}=M.  \label{th4}
\end{equation}
Eq.(\ref{th3}) and (\ref{th4}) imply that $S(\sigma _A^{\otimes n}),S(\sigma
_B^{\otimes n})\leq S(\sigma _{AB}^{\otimes n})$ and $\lambda (\sigma
_A^{\otimes n}),\lambda (\sigma _B^{\otimes n})\succ \lambda (\sigma
_{AB}^{\otimes n}),$ completing the proof.

As shown in the Ref.\ref{10}, attempts to characterize separability based
only upon studying the local and global properties of the eigenvalue spectra
of the system $\lambda (\sigma _A^{\otimes n}),\lambda (\sigma _B^{\otimes
n})$ and $\lambda (\sigma _{AB}^{\otimes n})$ can never work. Theorem 2
shows that the LPBS criterion is more close to the principal character of a
separable state than the inorder or entropic criterions. This means that the
number of product vectors of a mixed state are relative to the nonlocality
as shown in Ref \ref{q,f} which showed the number of product vectors of
states affects the local distinguishability of the states.

Although Theorem 1 is not sufficient for separability, we can get following
result:

Theorem 3: For the mixed state $\sigma $ in Eq.(\ref{a}) if the dimensions
of the LPBS is $2^{nS(\sigma )}$ and there exists a set of orthogonal
product bases of the LPBS, then $\sigma $ is separable.

Proof: If the dimensions of the LPBS of $\sigma ^{\otimes n}$ is $%
2^{nS(\sigma )},$ then $2^{nS(\sigma )}$ ``likely'' strings $\left|
string_j\right\rangle s$ are the linear compositions of $2^{nS(\sigma )}$
normalized product states $\left| ps_l^{\prime }\right\rangle s$ of the
LPBS, i.e.,

\begin{equation}
\left| string_j\right\rangle =A_{jl}^{\prime }\left| ps_l^{\prime
}\right\rangle ,\qquad j,l=1,\cdots ,2^{nS(\sigma )}.  \label{th5}
\end{equation}
$\left| string_j\right\rangle s$ and $\left| ps_l^{\prime }\right\rangle s$
are two set of bases of the LPBS. $\left| ps_l^{\prime }\right\rangle s$ are
orthogonal means that the two set of orthogonal and normalized bases $\left|
string_j\right\rangle s$ and $\left| ps_l^{\prime }\right\rangle s$ are
connected by matrix $A_{jl}^{\prime }.$ Linear algebra means that the matrix 
$A_{jl}^{\prime }$ is unitary. After two sides of Eq.(\ref{th5}) is
multiplied by $p=\prod_{i=1}^m\lambda _i^{n\lambda _i}$ one can see the
product states$\sqrt{p}\left| ps_{l^{\prime }}\right\rangle s$ is a set of
pure states decomposition of the mixed state $\sigma .$ This ends the proof.

By theorem 1, we can gain a interesting result, which also can be followed
from Ref \ref{12}.

Result: For the mixed state $\sigma $ in Eq.(\ref{a}) if

\[
\overline{E}(\sigma )\equiv \sum_i\lambda _iE_i>S(\sigma ), 
\]
where $E_i$ is the entanglement of pure state $\left| \Phi _i\right\rangle ,$
$\sigma $ is an entangled state.

Proof: The proof of result 1 is easy. Since the dimensions of LPBS of $%
\sigma ^{\otimes n}$ is not less than the Schmidt numbers of a ``likely''
strings, $2^{n\overline{E}(\sigma )},$ if $\overline{E}(\sigma )>S(\sigma ),$
then $\sigma $ is inseparable by Theorem 1, completing the proof.

The dimensions of LPBS of $\sigma ^{\otimes n}$ can be expressed as:

\[
\dim _{LPBS}(\sigma ^{\otimes n})=2^{n\overline{E}(\sigma )}.2^{nS^{\prime
}(\sigma )}, 
\]
where $2^{n\overline{E}(\sigma )}$ is equal to the least numbers of product
bases of a ``likely'' strings; $2^{nS^{\prime }(\sigma )}$ denotes the extra
numbers of product bases of LPBS owing to the cases where some different
``likely'' strings have different product bases. $0\leq S^{\prime }(\sigma
)\leq S(\sigma ).$ $S^{\prime }(\sigma )=0$ means that all ``likely''
strings have same product bases. $S^{\prime }(\sigma )=S(\sigma )$ means
that linearly independent product bases of each ``likely'' strings are
linearly independent to those of the other ``likely'' strings.

For example: $\sigma =\lambda \left| \Phi ^{+}\right\rangle \left\langle
\Phi ^{+}\right| +(1-\lambda )\left| \Phi ^{-}\right\rangle \left\langle
\Phi ^{-}\right| ,\left| \Phi ^{\pm }\right\rangle =\frac 1{\sqrt{2}}(\left|
00\right\rangle \pm \left| 11\right\rangle )$ (the mixture of any two Bell
states can be expressed as this form). $\left| \Phi ^{\pm }\right\rangle $
have same Schmidt bases, and then all ``likely'' strings of $\sigma
^{\otimes n}$ have same Schmidt bases. Thus $S^{\prime }(\sigma )=0,%
\overline{E}(\sigma )=1.$ So if $\sigma $ is separable, $S(\sigma )=1.$

In summary, we have gained a geometric diagram of a separable state $\sigma $%
: There exist $2^{nS(\sigma )}$ linearly independent product vectors which
span a same Hilbert space as $2^{nS(\sigma )}$ ``likely'' strings of $\sigma
^{\otimes n}$ do, i.e., the dimensions of LPBS of $\sigma ^{\otimes n}$ is $%
2^{nS(\sigma )}$. This diagram results in a criterion for separability,
i.e., product bases space criterion, which is strictly stronger than the
inorder criterion in \cite{12}. The product bases space criterion means that
the separability of a mixed state has close link to information entropy (or
the loss of classical information) and the number of product bases of the
mixed state. Indeed, the inorder criterion and the entropic criterion have
shown the connection between the separability and information entropy; Ref. 
\cite{15,16} have shown the connection between the separability and product
bases. The properties of product bases of a mixed state is also relative to
the local distinguishability of orthogonal states \cite{f}. Understanding
the relationship of the properties of product bases and nonlocality may be
an interesting problem for further research.

Chen would like to thank professor Guangcan Guo for his help to this work.

\end{document}